\documentclass[useAMS,usenatbib]{mn2e}

\usepackage {graphicx}
\usepackage{captcont}
\usepackage{times}
\def\aj{\,{AJ}}
\def\apj{\,{\rm ApJ}}
\def\apjl{\,{\rm ApJL}}

\def\aap{\,{\rm A\&A}}
\def\nat{\,{\rm nature}}
\def\mnras{\,{\rm MNRAS}}

\title[Are Passive Spiral Galaxies Truly ``Passive'' and ``Spiral''?]
{Are Passive Spiral Galaxies Truly ``Passive'' and ``Spiral''? :
Near-Infrared Perspective}
\author[Yamauchi \& Goto]
{
  Chisato Yamauchi$^{1,2}$\thanks{E-mail:cyamauch@a.phys.nagoya-u.ac.jp}
  and Tomotsugu Goto$^{3}$\thanks{E-mail:tomo@jhu.edu}
  \\
  $^{1}$Department of Physics and Astrophysics, Nagoya University,
  Chikusa-ku, Nagoya 464-8602, Japan\\
  $^{2}$National Astronomical Observatory, 2-21-1 Osawa, Mitaka, Tokyo
  181-8588, Japan\\
  $^{3}$ Department of Physics and Astronomy, The Johns Hopkins
  University, 3400 North Charles Street, Baltimore, MD 21218-2686, USA
}
\begin{document}


\pagerange{\pageref{firstpage}--\pageref{lastpage}} \pubyear{2004}

\maketitle

\label{firstpage}

\begin{abstract}
  Passive spiral galaxies --- unusual galaxies with spiral morphology
 without any
 sign of on-going star formation --- have recently been discovered to
  exist preferentially in cluster infalling regions (at about the virial radius, or at a local galaxy density of $\sim 1$ Mpc$^{-2}$).  The
 discovery directly connects the passive spiral galaxies to the cluster
 galaxy evolution studies such as the Butcher-Oemler effect or the
 morphology-density relation, i.e., passive spiral galaxies are likely
 to be transition objects between high-z blue, spiral galaxies and
 low-z red, cluster early-type galaxies. Thus,  detailed study of
 passive spiral galaxies potentially could bring a new
 insight on the underlying physical mechanisms governing cluster galaxy
 evolution. However, in previous work, passive spiral galaxies are
 selected from the low resolution optical images with $\sim 1.5$ arcsec of
 seeing. Therefore, passive spirals could be a mis-identification of S0
 galaxies; or dusty-starburst galaxies which are not passive at all. 

 To answer these questions, we performed a deep, high-resolution, near-infrared imaging of
 32 passive spiral galaxies. Our high resolution $K$ band images show
 clear spiral arm structures. Thus, passive spirals are not
 S0s. Optical-infrared colour does not show any signs of
 dusty-starburst at all. Therefore, it is likely that they are truly
 ``passive'' and ``spiral'' galaxies in the midst of cluster galaxy
 evolution.

\end{abstract}


\section{Introduction}\label{intro}

\begin{figure}
\includegraphics[scale=0.4]{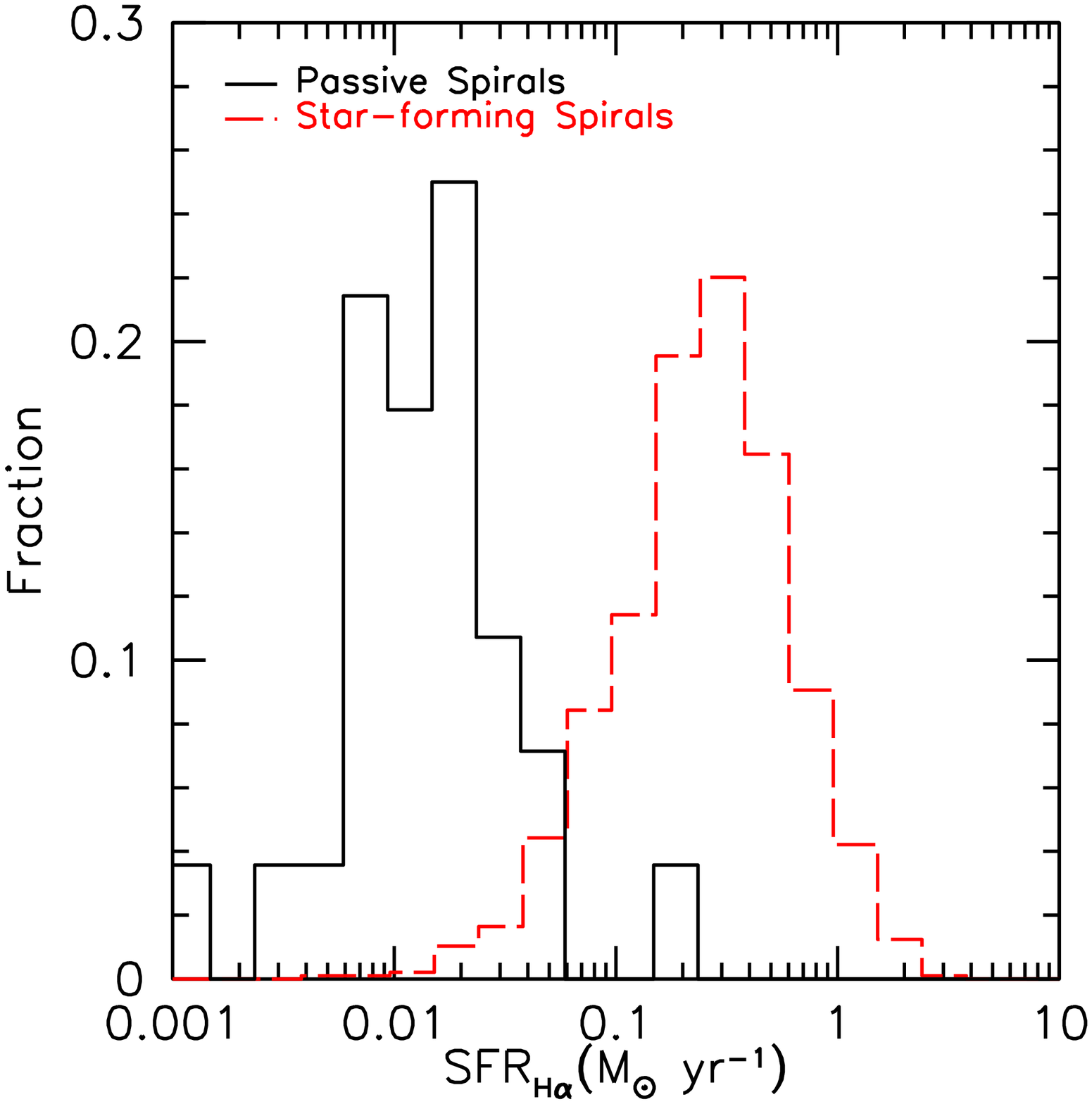}
\caption{The distributions of SFR estimated from the H$\alpha$ luminosity. The solid line and the dashed line are for the passive spiral galaxies and the star-forming galaxies, respectively. Note that these SFRs of passive spiral galaxies (the solid line) should be considered as upper limits since the H$\alpha$ line is not detected in these passive spiral galaxies and the continuum around H$\alpha$ wavelength is used to estimate the presented SFR.
}\label{fig:sfr_hist} 
\end{figure}

%
%
  
\begin{figure*}
\vspace*{20pt}
\includegraphics[scale=0.9]{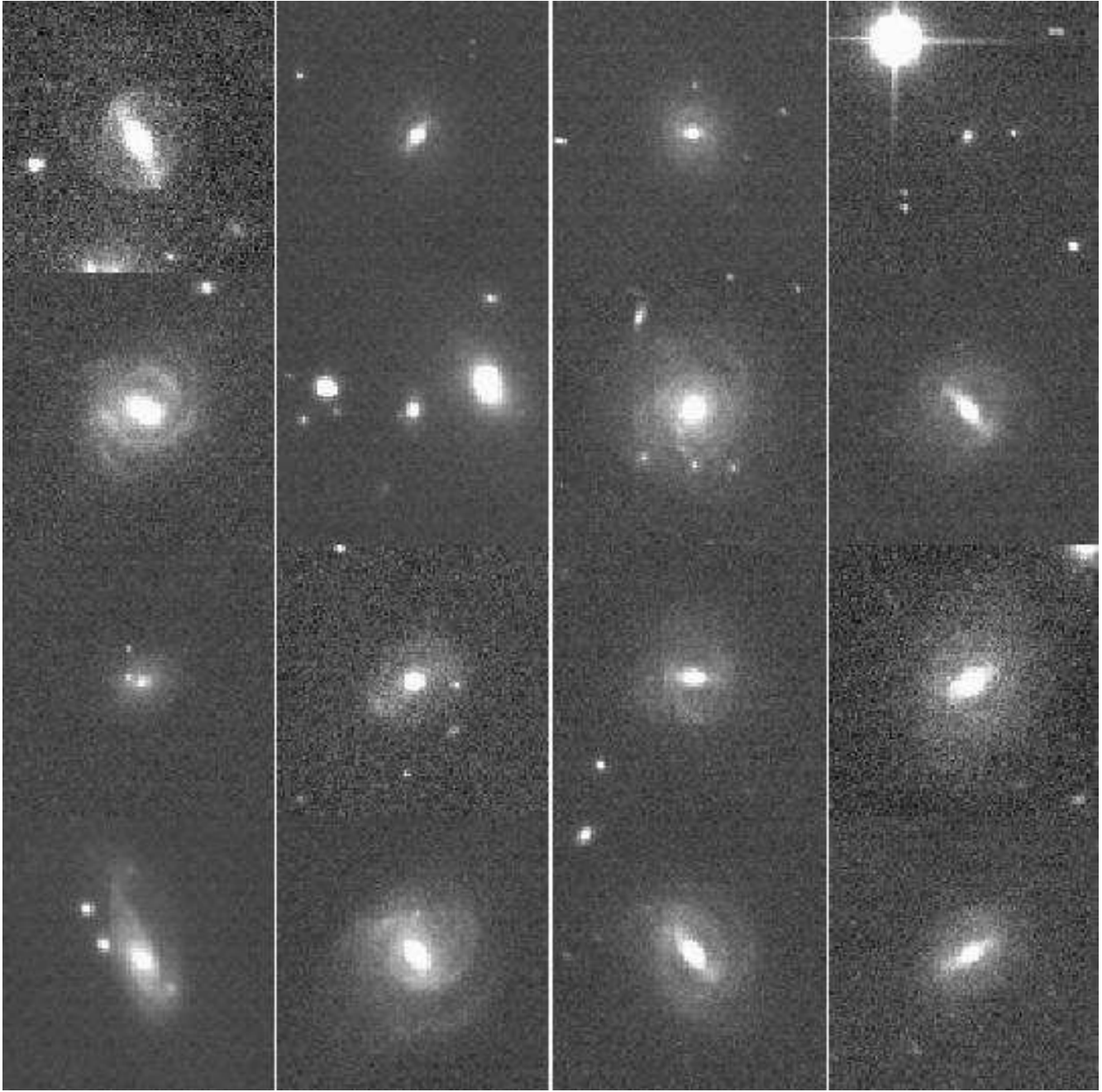}
\caption{
 The UKIRT $K$ band images of the observed passive spiral galaxies. Each image is
 35 $\times$ 35 arcsec. The figures are listed in the same order as in Table \ref{tab:targets} from the top left corner.
}\label{fig:k_image} 
\end{figure*}

\begin{figure*}
\vspace*{20pt}
\includegraphics[scale=0.9]{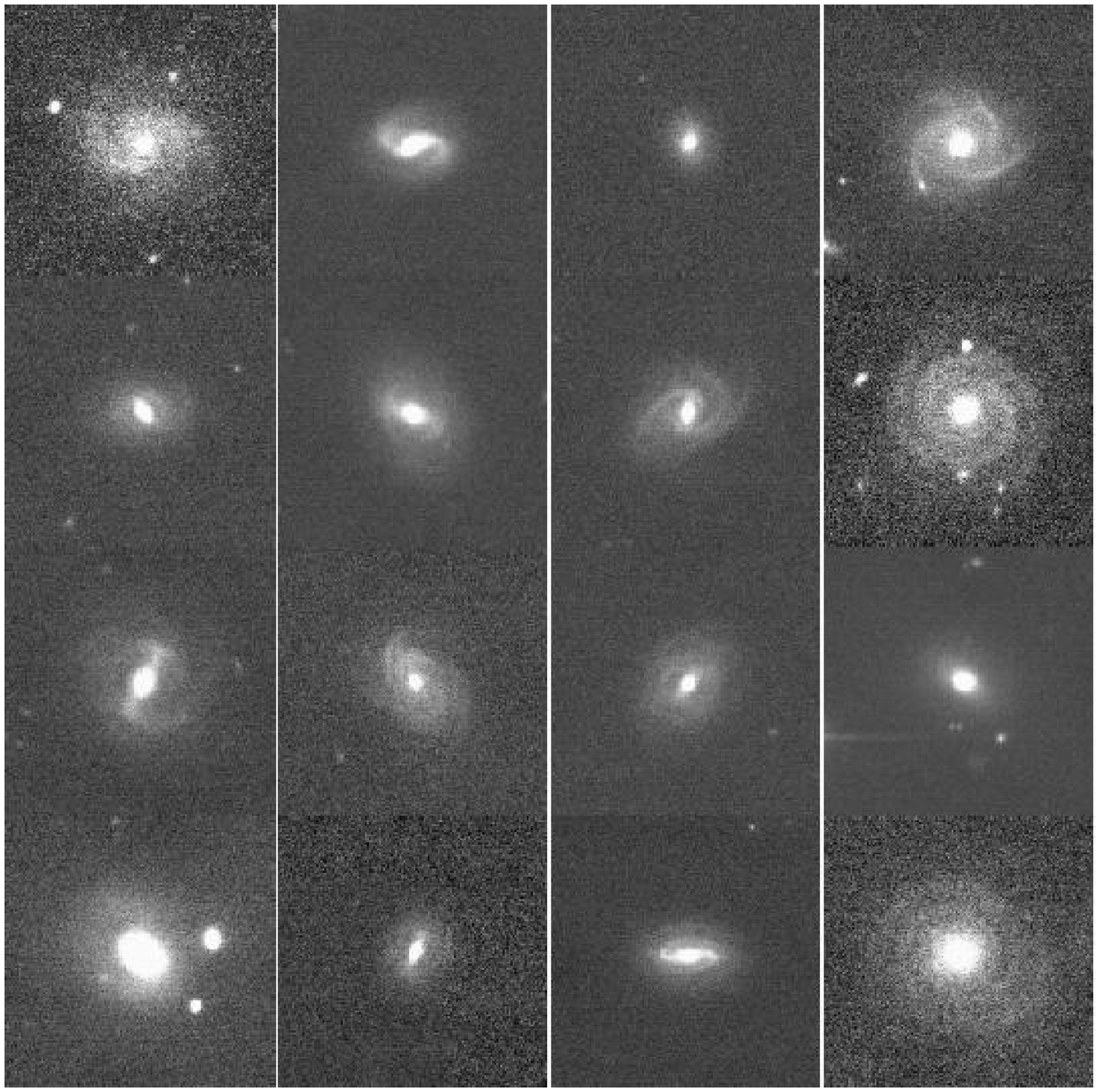}\\
\contcaption{}
\end{figure*}

   It is a remarkable feature that galaxy properties correlate with the
 environment where that galaxy exists. It has been well established that
 in the dense regions such as galaxy cluster cores, E/S0 galaxies are
 dominant, and that in the rarefied field regions, spiral galaxies are more
 numerous (Dressler 1980; Postman \& Geller 1984;
 Whitmore et al. 1993; Whitmore 1995; Dressler et
 al. 1997;  Hashimoto \& Oemler 1999; Fasano et al. 2000; Tran et
 al. 2001; Dom{\'{\i}}nguez et al. 2001, 2002; Helsdon \& Ponman
 2003; Treu et al. 2003; Goto et al. 2003a). This is the so-called
 morphology-density relation. As wide area CCD based
 surveys and large \& uniform galaxy cluster catalogs become available 
 (Postman et al. 1996; Annis et al. 1999; Kim et al. 2002; Goto et al. 2002a,b; Postman et
 al. 2002; Gal et al. 2003; Popesso et al. 2004), recent studies on the morphology-density relation
 started to reveal the environment where galaxy
 morphology start to change (Hogg et al. 2003; Blanton et al. 2003a;Tanaka et al. 2004). Goto et al. (2003a)  revealed that the morphology-density relation has two different breaks at local galaxy densities of
 1 Mpc$^{-2}$ and 0.3 Mpc$^{-2}$, possibly indicating the existence of
 two different physical mechanisms.

   Not only morphology but the star formation rate (SFR) of galaxies
 correlates with environments. It has been known for a long time that
 galaxy SFR is lower in the cluster core regions,
 resulting in numerous red galaxies in cluster cores (e.g., Couch \&
 Sharples 1987;Couch et al. 1994,1998; Dressler et al. 1994; Abraham et
 al. 1996; Pimbblet et al. 2002).  Recently it has become possible to
 specify the environment where SFR suddenly start to change (e.g. Kodama
 et al. 2001; Tanaka et al. 2004). Interestingly, this environment where SFR changes coincides
 with  the environment where galaxy morphology changes (Goto et
 al. 2004; Tanaka et al. 2004). Cluster galaxies change their SFR in the
 same environment where they change their morphology.

  It has also been known that cluster galaxies evolve. Butcher \& Oemler
  (1978,1984) found that fractions of blue galaxies in clusters increase
  with increasing redshift, i.e., cluster galaxies evolve from blue to
  red. This Butcher-Oemler effect was later
  confirmed by many authors (Rakos, Schombert 1995; Couch et al. 1994,1998;
  Margoniner, de Carvalho 2000; Margoniner et al. 2001; Ellingson  et
  al. 2001; Kodama \& Bower 2001; Goto et al. 2003b, but also see
  Andreon et al. 1999,2003). High redshift clusters ($z\sim 0.9$) are also known to
  have larger fractions of star-forming galaxies than local clusters
  (Postman, Lubin, \& Oke 1998; Postman, Lubin, \& Oke 2001). 
   Morphologically, it is  found that fractions of S0 galaxies are higher
  in high redshift clusters (Dressler et al. 1997; van Dokkum et al. 1998; Fasano et al. 2000; Jones, Smail, \& 
Couch 2000;    Fabricant et al. 2000;  also see Andreon et al. 1998).
  This claim was confirmed later by Goto et al. (2003b,2004) using the statistical number of 516  clusters found in the Sloan Digital Sky Survey (SDSS; Goto et al. 2002a,b).

    From these numerous pieces of observational evidence, we know that some
   physical mechanism is changing the morphology and SFR of cluster
   galaxies as a function of  the redshift. 
   However to date, it has been difficult to specify what physical mechanisms determine
    morphology and SFR of galaxies. It has been simply difficult to
 trace the complicated process of galaxy evolution with several Giga
 years of timescale, using the observation of only a single epoch.

 However, recently, a population of galaxies which are likely to shed
 some light on the subject has been actively debated. 
 The galaxies are called passive spiral galaxies (Couch et al. 1998;
 Dressler et al. 1999; Poggianti et al. 1999). Despite their
 spiral appearances, passive spirals do not have any
 emission lines indicative of on-going star formation.
 Passive spirals have been known to exist in many cluster studies (van
 den Bergh 1976; Wilkerson 1980; Bothun \& Sullivan 1980; Phillipps 1988; Cayatte et 
al. 1994; Couch et al. 1998; Poggianti et al. 1999; Bravo-Alfaro et 
al. 2001; Elmegreen et al. 2002). However, their
 abundance in the field region was not studied well. And thus, their
 connection to the cluster regions has not been clear until the
 recent discovery by Goto et al. (2003c), which claims that passive spiral
 galaxies exist preferentially in perimeter regions of galaxy clusters
 at around the virial radius of or local galaxy density of $\sim
 1$ Mpc$^{-2}$.  
Suggesting that passive spiral galaxies are created by some cluster related physical
mechanism, this discovery will bring significant implications on the underlying
physical mechanism.
Since passive spirals are expected to evolve into red, early type
cluster galaxies in a few Gyr, passive spirals are likely to be intermediate
transition objects between high-z blue, spirals and low-z red, early
type cluster galaxies. And thus, by studying passive spiral galaxies in
detail, we may be able to specify the physical mechanism responsible for
the cluster galaxy evolution.

 However, since the result of Goto et al. (2003c) was based on the SDSS and the Two Micron All Sky Survey  (2MASS; Jarrett et al. 2000) data, which have relatively poor image resolution ($\sim 1.5$
  arcsec of seeing) and large photometric error,  
 there have been two remaining important uncertainties 
  before interpreting passive spirals as transition objects:\\
 (i) Are passive spirals not S0s?; \\
 (ii) Are passive spirals not dusty starburst galaxies?\\
 If Goto et al. (2003c) mis-identified some S0 galaxies as passive
  spirals due to the poor seeing condition of the SDSS ($\sim 1.5$ arcsec), their
  discovery is less interesting since S0 galaxies are more common and
  very well studied in the literature. Also, if the emission lines of passive
  spirals are just suppressed by the heavy obscuration by dust
  in optical wavelength, passive spirals might not be passive at all. 

 In order to answer these two questions, we performed a deep $K$ band
 imaging of passive spiral galaxies. Deeper imaging will reveal detailed
 morphology of passive spirals. In addition, near-infrared light is less
 affected by the dust extinction, and thus allows us to distinguish
 dusty starburst galaxies from passive spirals.

 This paper is organized as follows: In
 Section \ref{data}, we describe the deep $K$ band observations we performed;
 In Section \ref{Nov 28 15:00:36 2003}, we present the results;
  In Section \ref{results}, we discuss the physical implications of our results;
 In Section \ref{summary}, we summarize our work and findings.
   The cosmological parameters adopted throughout this paper are $H_0$=75 km
 s$^{-1}$ Mpc$^{-1}$, and
($\Omega_m$,$\Omega_{\Lambda}$,$\Omega_k$)=(0.3,0.7,0.0).

%

\section{UKIRT Observation}\label{data}
 
 We have selected our target galaxies from 73 passive spiral galaxies presented in
 Goto et al. (2003c). All 73 passive spiral galaxies do not have any
 emission in [OII] nor H$\alpha$ ($< 1 \sigma$ in equivalent width) and have disc-like morphology.
 This sample is selected from a volume limited sample of galaxies ($0.05 < z < 0.1$, $M_r<-20.5$) based on the Sloan Digital Sky Survey data (Abazajian et al. 2003), and therefore is free from Malmquist type of bias.
 Fig. \ref{fig:sfr_hist} shows the distribution of star formation rate (SFR) for these 73 passive spiral galaxies computed from the luminosity in H$\alpha$ (the solid line).  The SFR is calculated using a conversion formula given in Kennicutt (1998), assuming constant extinction of 1 magnitude at the wavelength of H$\alpha$. We caution readers that these SFRs of passive spiral galaxies should be considered as upper limits since the H$\alpha$ line is not detected in these passive spiral galaxies and the continuum around H$\alpha$ wavelength is used to estimate the SFR in Fig.  \ref{fig:sfr_hist}.
    As a comparison sample, we have selected star-forming spiral galaxies as galaxies with both detected [OII] and H$\alpha$ emission lines (with $>1\sigma$ significance) and with the concentration index consistent to be a spiral ($Cin_r>0.5$). Here, $Cin_r$ is defined as the ratio of Petrosian 50\% flux radius to  Petrosian 90\% flux radius in $r$ (Shimasaku et al. 2001).
The distribution of SFR in the star-forming spiral galaxies is shown with the dashed line in Fig. \ref{fig:sfr_hist}. Compared with the star-forming spiral galaxies (the dashed line), our target passive spiral galaxies (the solid line) have lower SFR by about an order (or more). The difference again demonstrate that our target passive spiral galaxies indeed have much lower SFR than normal spiral galaxies in the field region.


 Among 73 passive spiral galaxies, all 32 passive spiral galaxies accessible during the run on
 2003 September 10-11 were observed on the 3.8m United Kingdom Infrared Telescope (UKIRT)
 in the $K$ band (2.2 $\mu$m) using the UKIRT Fast Track Imager (UFTI). 
 UFTI is a 1-2.5 $\mu$m InSb imager with 1024 $\times$ 1024 pixel.
 Each pixel subtends 0.091 arcsec on the sky. UFTI's field of view is
 $\sim 90$ arcsec.  
 Data were taken in periods of good atmospheric
 transparency and with the excellent seeing of $\sim$0.5 arcsec.  For
 each galaxy, we used two sets of 3$\times$3 grid of dithered position exposures of 60 sec each. The
 integration time is 18 mins for each galaxy. Calibration was obtained
 by observing a selection of UKIRT faint standards (Hawarden et al. 2001) on each night.
 The data were analyzed with the ORAC-DR data reduction pipeline which
 automates the dark subtraction, flat-fielding, re-sampling and de-spiking
 processes. 
 The photometry was performed using IRAF (v2.12.1) {\tt phot}    
  routines within 2$\times$ Petrosian radius measured in the SDSS $r$ band image (Stoughton et al. 2002).

 In addition to the 32 passive spiral galaxies, we have observed 12
 early-type galaxies as a control sample. These early-type galaxies were
 selected  from the same volume limited sample ($0.05 < z < 0.1$,
 $M_r<-20.5$)  as galaxies with $Cin_r<0.4$, and thus mainly consist of
 elliptical and S0 galaxies. We compare this control sample with passive
 spiral galaxies in order to test truly ``passive'' and ``spiral''
 nature of the galaxies. 

 The purpose in using $K$ band is two-folded. First, the $K$ band is
 relatively free from dust extinction. Therefore, we can test dust
 extinction in passive spirals using $r-K$ colour. Also, since the $K$ band
 traces the old stellar population, i.e., the dominant mass
 distribution in galaxies, it is suitable to study galaxy morphology.

\begin{table*}
\begin{center}
\caption{
List of observed targets.
}\label{tab:targets}
\begin{tabular}{lrrrrrrl}
\hline
Name                     & R.A.         &  Dec.    & Redshift  & $K$   & $K_{err}$& $C_{in}$(K) & \\
\hline
\hline
SDSSJ004339.22+151025.6  &   0:43:39.22 &  15:10:25.64 & 0.081 & 13.789 & 0.015 & 0.26 &  \\
SDSSJ010647.55+140048.1  &   1:06:47.55 &  14:00:48.13 & 0.089 & 13.532 & 0.005 & 0.39 &  \\
SDSSJ010955.90+154757.4  &   1:09:55.90 &  15:47:57.40 & 0.062 & 13.344 & 0.007 & 0.47 &  \\
SDSSJ012409.19-002555.9  &   1:24:09.19 &  -0:25:55.97 & 0.080 & 16.032 & 0.025 & 0.14 &  $\dagger$ \\
SDSSJ012528.30+004411.7  &   1:25:28.30 &   0:44:11.75 & 0.089 & 12.513 & 0.006 & 0.47 &  \\
SDSSJ015855.15-095143.2  &   1:58:55.15 &  -9:51:43.25 & 0.082 & 14.316 & 0.005 & 0.93 &  $\ddagger$ \\
SDSSJ021534.36-090537.0  &   2:15:34.36 &  -9:05:37.06 & 0.069 & 12.219 & 0.004 & 0.49 &  \\
SDSSJ024732.02-065137.4  &   2:47:32.02 &  -6:51:37.48 & 0.071 & 13.044 & 0.008 & 0.49 &  \\
SDSSJ033322.66-000907.5  &   3:33:22.66 &  -0:09:07.51 & 0.085 & 14.178 & 0.019 & 0.52 &  \\
SDSSJ074452.51+373852.7  &   7:44:52.51 &  37:38:52.73 & 0.074 & 13.932 & 0.008 & 0.49 &  \\
SDSSJ143320.16+003952.7  &  14:33:20.16 &   0:39:52.71 & 0.078 & 12.890 & 0.010 & 0.53 &  \\
SDSSJ151033.69+021434.8  &  15:10:33.69 &   2:14:34.81 & 0.074 & 12.821 & 0.013 & 0.30 &  \\
SDSSJ151747.79+030052.1  &  15:17:47.79 &   3:00:52.15 & 0.082 & 11.606 & 0.004 & 0.46 &  \\
SDSSJ152621.67+035002.4  &  15:26:21.67 &   3:50:02.46 & 0.083 & 12.017 & 0.006 & 0.46 &  \\
SDSSJ161125.21+524526.8  &  16:11:25.21 &  52:45:26.87 & 0.063 & 12.524 & 0.007 & 0.49 &  \\
SDSSJ161655.51+521449.2  &  16:16:55.51 &  52:14:49.25 & 0.089 & 13.559 & 0.012 & 0.53 &  \\
SDSSJ163340.30+475018.4  &  16:33:40.30 &  47:50:18.44 & 0.061 & 13.496 & 0.010 & 0.57 &  \\
SDSSJ174218.49+551537.5  &  17:42:18.49 &  55:15:37.53 & 0.062 & 12.973 & 0.010 & 0.53 &  \\
SDSSJ222206.46-011002.7  &  22:22:06.46 &  -1:10:02.77 & 0.100 & 14.402 & 0.016 & 0.48 &  \\
SDSSJ223239.13-082323.1  &  22:32:39.13 &  -8:23:23.13 & 0.080 & 12.863 & 0.008 & 0.54 &  \\
SDSSJ223558.39-002313.8  &  22:35:58.39 &  -0:23:13.84 & 0.080 & 13.660 & 0.013 & 0.48 &  \\
SDSSJ224000.16-004945.1  &  22:40:00.16 &  -0:49:45.15 & 0.053 & 12.683 & 0.009 & 0.51 &  \\
SDSSJ224435.97-081615.5  &  22:44:35.97 &  -8:16:15.58 & 0.082 & 13.047 & 0.014 & 0.50 &  \\
SDSSJ224747.17+125125.9  &  22:47:47.17 &  12:51:25.96 & 0.092 & 12.990 & 0.009 & 0.59 &  \\
SDSSJ231642.88+153954.5  &  23:16:42.88 &  15:39:54.58 & 0.091 & 12.882 & 0.010 & 0.52 &  \\
SDSSJ232259.44+145915.4  &  23:22:59.44 &  14:59:15.46 & 0.095 & 13.570 & 0.012 & 0.56 &  \\
SDSSJ233354.88-002449.3  &  23:33:54.88 &  -0:24:49.39 & 0.088 & 13.214 & 0.008 & 0.54 &  \\
SDSSJ234036.90+142943.4  &  23:40:36.90 &  14:29:43.44 & 0.067 & 12.877 & 0.003 & 0.41 &  \\
SDSSJ234206.04+150129.0  &  23:42:06.04 &  15:01:29.05 & 0.066 & 12.056 & 0.006 & 0.44 &  \\
SDSSJ234523.98+010552.0  &  23:45:23.98 &   1:05:52.00 & 0.059 & 13.787 & 0.028 & 0.53 &  \\
SDSSJ235307.15+150355.3  &  23:53:07.15 &  15:03:55.38 & 0.079 & 13.174 & 0.013 & 0.46 &  \\
SDSSJ235741.11+004135.7  &  23:57:41.11 &   0:41:35.74 & 0.061 & 13.337 & 0.027 & 0.48 &  \\
\hline
\end{tabular} 
\end{center}
\begin{flushleft}
$\dagger$ -- A bright star in the same field.\\
$\ddagger$ -- A possible overlap with a bright nearby galaxy. \\
\end{flushleft}
\end{table*}
\section{Results}

\label{Nov 28 15:00:36 2003}
\subsection{K-band Galaxy Morphology}\label{image}

\begin{figure}
\includegraphics[scale=0.4]{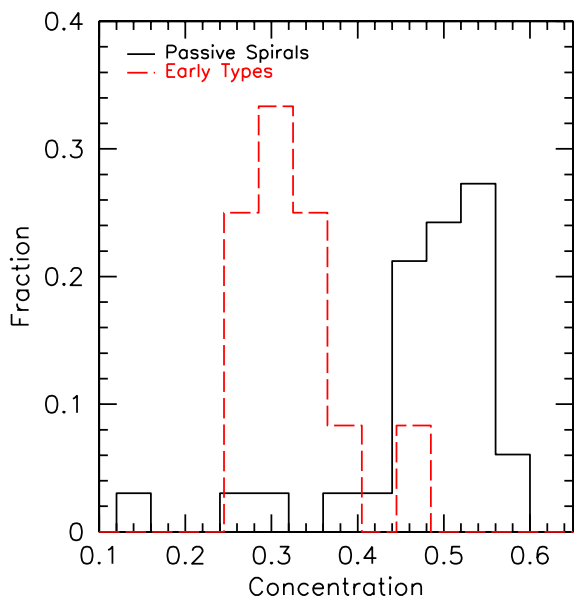}
\caption{
 Distribution of concentration parameter, $Cin_K$, measured as a ratio of
 Petrosian 50 to 90\% flux radius in $K$ band. 
}\label{fig:kconcent} 
\end{figure}

   First, we present $K$ band morphology of passive spirals to make sure
 that they are not S0 galaxies, which are more common and known to have passive nature.
 In Fig. \ref{fig:k_image}, we show $K$ band images of 32
 passive spiral galaxies taken with the UKIRT. 
 The deep and high resolution imaging capability of UKIRT can probe the discs and spiral arm structures in the passive spiral galaxies. 
 Although there are two cases where disc structures are not clear due to the presence of a nearby bright object (SDSSJ012409.19-002555.97 and SDSSJ015855.17-105143.25), 
 the rest of the passive spiral galaxies have discs and spiral arm structures
 without contamination from more common S0 galaxies. 

  Qualitatively, we measured the concentration of passive spiral
 galaxies as a ratio of Petrosian 50\% flux radius to 90\% flux radius
 using the flux measured in the $K$ band.  Note that this concentration
 parameter, $Cin_K$, is an inverse of the commonly used concentration
 parameter, and thus, later-type galaxies have $larger$ values of
 $Cin_K$. Since our $K$ band images are much deeper with twice as high
 resolution as the SDSS images,  $Cin_K$ provides us with a better description of
 galaxy morphology than the concentration parameter used in Goto et
 al. (2003).  Since the seeing size was almost constant during the two
 days of observation ($\sim 0.5$ arcsec) and our galaxies are at a
 similar redshift ($z \sim 0.08$), we did not correct for the seeing.
 In Fig. \ref{fig:kconcent}, we show the distribution of $Cin_K$ for
 both the passive spirals (the solid line) and the control sample (the dashed
 line). Reassuringly, passive spiral galaxies and the control sample of
 early type galaxies have a very different $Cin_K$ distribution, with
 passive spirals having much higher values of $Cin_K$. A
 Kolomogorov-Smirnov test shows that these two distributions are
 different with more than 99.99\% significance. This difference in the $Cin_K$
 distribution assures that our passive spiral galaxies are indeed
 different galaxy population than well-studied S0 galaxies.

\subsection{Optical-Infrared Colour}\label{color}

\begin{figure}
\includegraphics[scale=0.4]{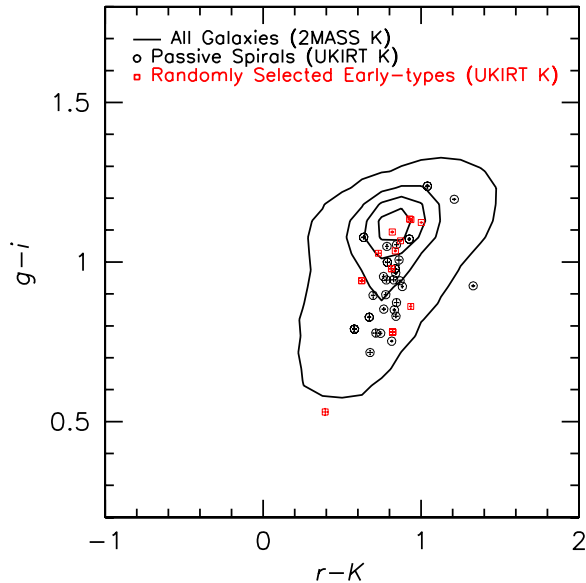}
\caption{
 Restframe $g-i$ vs. $r-K$ two-colour diagram. The circles are for
 passive spirals. The squares are for the early-type galaxies in the control sample. The
 contours represent all galaxies in the volume limited sample with 2MASS
 $K$ magnitude. The error bars are shown inside the circles and squares.
}\label{fig:rk} 
\end{figure}

\begin{figure}
\includegraphics[scale=0.4]{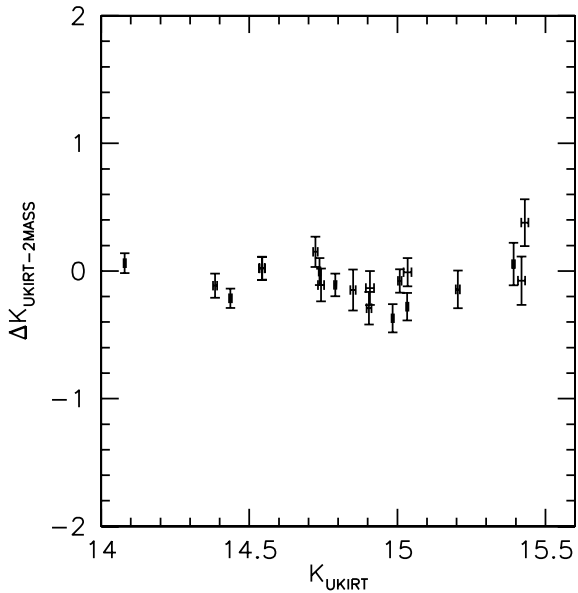}
\caption{
 Offsets between UKIRT $K$ magnitude and 2MASS $K$ magnitude are shown
 for 22 galaxies commonly observed with both of the two telescopes. The
 mean deviation and rms is 0.05 and 0.17, respectively. Note that $K$ is in AB system in this figure.
}\label{fig:offset} 
\end{figure}

 Next, we present optical-infrared ($r-K$) colour distribution of
 passive spiral galaxies in order to check whether they are
 dusty starburst galaxies or truly passive galaxies. Since $K$ band is
 less affected by the dust extinction than $r$ band, dusty starburst galaxies
 are known to have redder colours in $r-K$ by $\sim 1$ mag (Smail et al. 1999) . 
  Fig. \ref{fig:rk} plots $g-i$ colour against $r-K$ colour. Optical
 photometries ($g,r,$ and $i$) are from the SDSS and $k$-corrected to
 the restframe using the routine given in Blanton et al. (2003b; v1\_11).
 The black circles are for passive spiral galaxies observed with UKIRT. The
 squares are for early-type galaxies in the control sample. For a reference,
 we plot the distribution of all galaxies in the volume limited sample
 with $K$ magnitudes measured with the Two Micron All Sky Survey (2MASS;
 Jarrett et al. 2000) as the contour. When comparing
 2MASS $K$ magnitude with the UKIRT $K$ magnitudes, we found a slight
 shift between these two magnitudes as shown in Fig. \ref{fig:offset}. 
 We have calibrated this offset using 22 galaxies commonly observed
 with both 2MASS and UKIRT to match UKIRT $K$ mag to 2MASS $K$ mag.
 $K$ band magnitudes are $K$-corrected using  Mannucci et
 al. (2001).
  The error bars are plotted as horizontal and perpendicular bars. Compared
 with the error with 2MASS $K$ ($\Delta K \sim 0.1$), the UKIRT observation reduced the error
 in $K$ magnitude significantly ($\Delta K < 0.03$; c.f. Fig. 13 of
 Goto et al. 2003c). 
  Interestingly, compared with all galaxies (the contour), passive
 spiral galaxies (circles) are not redder at all in $r-K$ colour. Indeed,
 $r-K$ colours of passive spiral galaxies are indistinguishable from
 early-type galaxies (squares). 
 These results support truly passive nature of these galaxies
 since dusty starburst galaxies should have $r-K$ colour by 1 magnitude redder than normal
 galaxies (i.e., $r-K \sim 1.8$; See Smail et al. 1999).

\section{Discussion}\label{results}

 In Section \ref{Nov 28 15:00:36 2003}, we have answered the two remaining
 questions on passive spiral galaxies using the deep $K$ band imaging
 with UKIRT; (i) passive spiral galaxies indeed have discs \& spiral arm
 structures, and therefore they are a different population of galaxies
 from S0 galaxies; (ii) optical-infrared ($r-K$) colour does not show
 any  signs of dusty starburst galaxies. Therefore, they must be truly
 passive galaxies.
   In this section, we discuss physical implications of our results. 
 Since our UKIRT observation has secured that passive spiral galaxies are
 truly spiral galaxies with no star formation, we now have to take it
 more seriously that these passive spiral galaxies exist in the cluster
 perimeter regions (Goto et al. 2003c). 

 It has been long discussed whether the properties of cluster galaxies are by their 'nature' or 'nurtured' later. According to the standard hierarchical clustering model, galaxies in high density regions of the Universe such as galaxy clusters have collapsed earlier, and thus more evolved than galaxies in the low density field regions. In addition to this, galaxies in dense regions have been subject to additional physical mechanisms specific to the dense regions. Therefore, it is important to understand whether properties of cluster galaxies were established early in the universe when the galaxy assembled (nature), or they are later formed by the physical mechanisms specific to the dense regions (nurture). Goto et al. (2003c) have found that passive spiral galaxies preferentially exist in cluster perimeter regions. If passive spiral galaxies do not exist in the low density field regions, the result may favor the 'nurture' scenario where only cluster specific physical mechanisms can create passive spiral galaxies.

   Various physical mechanisms have been proposed to explain the
 cluster galaxy evolution. Possible mechanisms include ram-pressure stripping
 of gas (Gunn \& Gott 1972; Farouki \& Shapiro 1980; Kent 1981; Fujita \& Nagashima 1999;
  Abadi, Moore \& Bower 1999; Quilis, Moore \& Bower 2000; Fujita \& Goto 2004);
  galaxy harassment via high speed impulsive
  encounters (Moore et al. 1996, 1999; Fujita 1998); cluster
 tidal forces (Byrd \& Valtonen 1990; Valluri 1993; Fujita 1998; Gnedin 2003a,b) which
 distort galaxies as they come close to the centre; interaction/merging of
 galaxies (Icke 1985; Lavery \& Henry 1988; Mamon 1992; Makino \& Hut
  1997; Bekki 1998;  Finoguenov et al. 2003a);  evaporation of the cold
 gas in disc galaxies via heat conduction from the surrounding hot ICM
 (Cowie \& Songaila 1977; Fujita 2003); and a gradual decline in
 the SFR of a galaxy due to the stripping of halo gas (strangulation or
 suffocation; Larson, Tinsley \& Caldwell 1980; Bekki et al. 2002;
 Kodama et al. 2001; Finoguenov et al. 2003b). 

   Among all of these, strong dynamical interactions such as  cluster
 tidal forces and major interaction/merging of
 galaxies are less preferred since such processes distort the morphology
 of galaxies and cannot explain spiral arm structures in passive spiral
 galaxies. 

  Among the rest, since passive spirals exist in the environment with
 a  local galaxy density of $\sim 1$ Mpc$^{-2}$ or at about the virial radius, those mechanisms that work in this environment are
 good candidates for the creation of passive spiral galaxies. Kodama et
 al. (2001) and Tanaka et al. (2004) discussed that plasma gas density
 is too low at the virial radius for ram-pressure stripping (of
 the cold gas in a galactic disc), concluding that stripping of hot halo
 gas is preferred to the stripping of cold gas. However, using the analytical model, Fujita (2003) showed that the cold gas stripping can be effective at around the virial radius at higher redshift (also see Fujita \& Goto 2004). Mihos et al. (2003) proposed that    infalling sub-groups may have high enough gas density to strip the cold gas.  

  Unfortunately, to further specify the responsible mechanisms is rather
 difficult. Any of the remaining mechanisms can work at the environment around the
 virial radius. 
 Since most of the processes act over a period of a few Gyr,
 observations at one redshift cannot easily provide 
 the detailed information that is needed to specify one process.
  It is also worth noting that  E+A (post-starburst) galaxies (Dressler \& Gunn 1983), which have
   been thought to be transition objects in cluster galaxy evolution,
   were found to have their origin in merger/interaction in the general
   field region (Goto et al. 2003d,e).  And thus, explaining cluster
   galaxy evolution using E+A galaxies is not realistic anymore.

 More importantly, Tanaka et al. (2004) found that the environmental
 dependence of galaxies properties is different for bright and faint
 galaxies. Faint galaxies ($M^*+1<M_r<M^*+2$) have a break at the same
 environment as this work. On the other hand, bright galaxies
 ($M_r<M^*+1$) do not have a specific break in environmental dependence,
 and their properties monotonically change as a function of the
 environment. Goto et al. (2003a) found two breaks on the
 morphology-density relation.
 These results might be indicating that there may be two (or more)
 different physical processes at work, and that passive spiral galaxies
 may be the transition objects of only one physical mechanism among many.
   
  However, discovering transition objects in a certain environment is one
  step forward compared with previous work. Since we now know what
  galaxies we should trace, observing the abundance and properties of
  passive spiral galaxies toward higher redshift clusters will bring
  further implications on the underlying physical mechanism. As a
  forerunner, Goto et al. (2004) identified a population of red,
  late-type galaxies rapidly changing morphology at $z \sim 0.17$. 
 Among  various semi-analytic simulations of cluster galaxy evolution  (e.g.,  Okamoto \& Nagashima 2001; Diaferio et al. 2001; Benson et al. 2001; Springel et al. 2001;
  Shioya et al. 2001,2002; Okamoto \& Nagashima 2003), 
  any that predict passive spirals as transition objects should be favored.


\section{Summary}\label{summary}

 We have performed a deep $K$ band imaging of 32 passive spiral galaxies
 with the UKIRT, in order to answer the remaining two questions in the
 subject: (i) passive spirals are S0s or not; 
 (ii) they are dusty starburst galaxies or not. Our results are summarized as
 follows.
 
 \begin{itemize}
  \item All 32 $K$ band images of passive spiral galaxies with seeing of
	$\sim 0.5$ arcsec show clear spiral arm structures in the disc, except two unclear cases due to a nearby bright object. The
	distribution of the concentration parameter is different from
	that of early-type galaxies with more than 99.99\%
	significance. We conclude that passive spirals are a different
	population of galaxies from S0s.
  \item Optical-infrared colour ($r-K$) of passive spiral galaxies is
	not redder than that of normal galaxies.  Therefore, passive
	spiral galaxies are not likely to be dusty starburst galaxies. 
 \end{itemize}

 Since our results support truly ``passive'' and ``spiral'' nature of
 these galaxies, it is very likely that passive spiral galaxies are
 indeed transition objects currently undergoing cluster galaxy
 evolution. Further study of passive spiral galaxies will have further implications for the physical mechanisms governing cluster galaxy evolution.

\section*{Acknowledgments}

We are grateful to Masayuki Tanaka and Sadanori Okamura for useful
discussion. We thank the referee, Dr. Philip James for many insightful comments, which improved the paper significantly.  
We thank the UKIRT support astronomers and telescope operators for their help during the observation.
The United Kingdom Infrared Telescope is operated by the Joint Astronomy
Centre on behalf of the U.K. Particle Physics and Astronomy Research
Council.


\label{lastpage}

\end{document}